\documentclass[10pt]{iopart}
\usepackage{graphicx}

\newcommand{\lf}[2]{\mbox{\Large $\frac{#1}{#2}$}}

\begin{document}

% Journal identifier to be inserted here: Classical and Quantum Gravity

\jl{6}

\title[Resonance widening\ldots]{Resonance widening in spherical GW
detectors: model descriptions of the dissipation processes}

\author{Eugenio Coccia\dag, Viviana Fafone\ddag, Giorgio Frossati\S,
Arlette de Waard\S, and {J.\ Alberto Lobo\P\footnote[3]{To whom
correspondence should be addressed.}}}
\address{\dag\ Dipartimento di Fisica, Universit\`a di Roma ``Tor Vergata'',
	{\rm and\/} \\ INFN Sezione di Roma 2, Via Ricerca 
	Scientifica 1, 00133 Roma, Italy}
\address{\ddag\ INFN Laboratori Nazionali di Frascati, Via E.\ Fermi 40,
	00044 Frascati (Roma), Italy}
\address{\S\ Kamerlingh Onnes Laboratory, Leiden University, Leiden, The
	Netherlands}
\address{\P\ Departament de F\'\i sica Fonamental,
	Diagonal 647, E-08028 Barcelona, Spain.}

\date{\today}
%\date{9 August 2002) \\ (Revised \today}

\begin{abstract}
Internal friction effects are responsible for line widening of the
resonance frequencies in mechanical oscillators and result in damped
oscillations of its eigenmodes with a decay time $Q/\omega\/$. We study
the solutions to the equations of motion for the case of spherical
oscillators, to be used as next generation of acoustic gravitational wave
detectors, based on various different assumptions about the material's
constituent equations. Quality factor dependence on mode frequency
is determined in each case, and a discussion of its applicability
to actual gravitational wave detectors is made on the basis of
available experimental evidence.
\end{abstract}


%\submitted

\section{Introduction}
\label{sec.1}

Spherical Gravitational Wave (GW) detectors will almost certainly be,
in one of its variants (solid, hollow or dual), the next generation of
acoustic antennae, due to their multimode-multifrequency capabilities
\cite{clo,jm,ls,mnras,nadja,vega,padua}, as well as their potentially
enhanced sensitivity relative to their currently operating cylindrical
counterparts~\cite{nau,as,padua2,alle,niobe}. Conviction that this is
going to be the case has encouraged a remarkable research effort within
the GW community, and a variety of important topics have been addressed,
ranging from theoretical to practical aspects of the problem. Several
countries worldwide are currently developing projects to build and
operate spherical detectors~\cite{omni,sfera,grail}.

Some of the salient properties of a spherical body as a detector
of GWs can be accurately established by means of an idealised model,
where dissipative effects can be safely overlooked~\cite{lobo,wp,bian}.
Such effects are however very important when it comes to determining
the ultimate sensitivity of the device, due to their fundamental
bearing on the {\it noise\/} characteristics of the antenna via the
fluctuation--dissipation theorem, more specifically its {\it spectral
density}.

In a solid elastic body there are many oscillation modes in addition
to the fundamental one chosen for GW detection, which therefore add
noise to the frequency band of operation. It has been suggested
\cite{levin,tung} that normal mode analysis of this contribution to the
noise level can be either inconvenient or too expensive in terms of
computational cost if heat is e.g.\ generated by a laser spot on the
oscillating solid, such as happens in an interferometric GW detector
or in a dual sphere~\cite{padua}. Experimental measurements clearly
show that this is the case~\cite{yama}.

The na\"\i ve approach, and actually the one most often taken, to
account for dissipative effects is to add an {\it ad hoc\/} term
proportional to the velocity in the solid's eigenfrequency equation.
This results in a damped oscillatory behaviour of the sort

\begin{equation}
        e^{-\omega t/Q}\,\sin\omega t   \label{I.1}
\end{equation}
where the {\it quality factor Q\/} is assumed constant (i.e., time
independent), and accounts for the linewidth of the mode with frequency
$\omega\/$. It is well known~\cite{cff} that this quality factor is
different for different oscillation modes, so the set of $Q\/$'s is
very large, and the procedure appears highly artificial. On the other
hand, there has been criticism of the concept of dissipation forces
proportional to velocity in terms of actual noise spectrum in an
elastic solid~\cite{majo}, and suggestion that a better description
may rather derive from the use of {\it complex\/} frequencies.

In the specialised literature on the subject (see~\cite{ja} and
references therein), viscoelastic effects are often described by
means of so called {\it constituent equations\/}. These are extensions
of Hooke's law relating the stress and strain tensors in the solid.
Much like in a simple one-dimensional spring, internal friction forces
can be considered proportional to the instantaneous velocity of the
oscillating mass, hence constituent equations usually contain time
derivatives of those tensors, and depend on a small number of viscosity
parameters, to be added to the elastic Lam\'e coefficients $\lambda$
and $\mu\/$ ---see below. There is however no unique way in which this
idea can be carried over from a one-dimensional system into a constituent
equations set for a three-dimensional solid, and so different alternatives
result in different models for the purpose.

In this paper we propose to study and discuss the results of applying to
a spherical GW detector the equations of various such phenomenological
models, in view to determine how quality factors change from mode to
mode in each case. This, it is hoped, will generate some insight into
the nature of the viscous processes which take place in a specific
spherical GW detector, and help to assess on the basis of spectral
measurements which particular class of viscoelastic solid a given
material belongs to. In turn, better understanding of material's
macroscopic properties should also contribute relevant information
to the currently important issue of spherical GW detector design. In
section~\ref{sec.2} we present the general equations, and in the subsequent
sections we successively consider the Kelvin--Voigt, Maxwell, Standard
Linear and Genaralised Mechanical models. As we shall see, the sphere's
vibration eigenmodes always group into the usual families of
{\it toroidal\/} (purely torsional) and {\it spheroidal\/} modes,
but different quality factor dependences arise in different models.
In section~\ref{sec.6} we match our theoretical results to a number
of experimental measurements of quality factors performed on various
spheres, and discuss which is the most appropriate model to fit such
data. In the final section we present a summary of conclusions, and
two appendices are added to clarify a few mathematical technicalities.

\section{The general equations}
\label{sec.2}

GWs bathing the Earth are known to be extremely weak ---see e.g. \cite{sh}
for a review--- so that the classical equations of linear Elasticity are
very good to describe the GW induced motions of a spherical antenna in
the expected frequency range, roughly 10$^2$--10$^4$ Hz. These equations
are \cite{ll}

\begin{equation}
    \rho\,\frac{\partial^2 s_i}{\partial t^2} -
    \frac{\partial\sigma_{ij}}{\partial x_j} = f_i  \label{eqmov}
\end{equation}
where ${\bf s}({\bf x},t)$ is the field of displacements in the elastic
body, and ${\bf f}({\bf x},t)$ is the GW induced density of forces acting
on the solid ---see \cite{lobo} for full technical details. $\sigma_{ij}\/$
is the the {\it stress\/} tensor, and is related to the {\it strain\/}
tensor

\begin{equation}
        s_{ij}\equiv\frac{1}{2}\,\left(\frac{\partial s_i}{\partial x_j}
               + \frac{\partial s_j}{\partial x_i}\right)    \label{I.2}
\end{equation}
through a set of {\it constituent equations}. In the case of a
non-dissipative (ideal) solid these are simply the expression of Hooke's
law. In a dissipative one, such equations include {\it time derivatives\/}
of both $s_{ij}\/$ and $\sigma_{ij}\/$ to account for internal friction
effects. Constituent equations are of the following general type:

\begin{equation}
   L(s_{ij},\dot s_{ij},\ddot s_{ij},\ldots;
     \sigma_{ij},\dot\sigma_{ij},\ddot\sigma_{ij},\ldots) = 0
   \label{I.3}
\end{equation}

In this paper we shall limit ourselves to {\it linear\/} constituent
equations, an excellent approach for a GW detector, as already stressed.
In the simplest instance, only first order derivatives will appear
in~\eref{I.3}, and we shall consider this first. Then we shall also
devote some attention to more complicted models.

The equations of motion~\eref{eqmov} must be supplemented with suitable
{\it boundary conditions\/}. We shall prescribe the usual ones

\begin{equation}
  \sigma_{ij}\,n_j = 0 \qquad {\rm at} \qquad r=R
  \label{I.4}
\end{equation}
where {\bf n} is a unit outward-pointing vector, expressing that the
surface of the sphere is free from any tensions and/or tractions.

\section{Kelvin--Voigt model}
\label{sec.3}

This model assumes that the solid is homogeneous and isotropic, and is
characterised by the following constituent equations \cite{ja}:

\begin{equation}
  \sigma_{ij}=\left(\lambda+\lambda'\frac{\partial}{\partial t}\right)
   s_{kk}\,\delta_{ij} + 2\left(\mu+\mu'\frac{\partial}{\partial t}\right)
   s_{ij}   \label{consKV}
\end{equation}

The constants $\lambda$ and $\mu$ are the usual Lam\'e coefficients
describing the purely elastic behaviour of the body \cite{ll}, while the
positive coefficients $\lambda'$ and $\mu'$ parametrise its {\it viscous\/}
properties\footnote{
These coefficients are actually analogous to those which describe the
viscosity of fluids in Hydrodynamics: {\it shear\/} viscosity ($\mu'$) and
{\it bulk\/} viscosity ($2\mu'+3\lambda'$).}, which are proportional to the
change rate of the strain tensor, $\partial_t s_{ij}$.

If equations~\eref{consKV} are replaced into~\eref{eqmov} we obtain the
equations of motion for ${\bf s}({\bf x},t)$:

\begin{equation}
  \hspace*{-1 cm}
  \rho\,\frac{\partial^2 {\bf s}}{\partial t^2} =
  \left(\mu + \mu'\frac{\partial}{\partial t}\right)\nabla^2 {\bf s} \,-
  \,\left[(\lambda+\mu)+(\lambda'+\mu')\frac{\partial}{\partial t}\right]
  \,\nabla(\nabla{\bf \cdot}{\bf s}) + {\bf f}({\bf x},t)
  \label{eqmovKV}
\end{equation}

The solution to this system of coupled equations can be expressed in terms
of a Green function, the construction of which requires explicit knowledge
of its {\it eigenmode\/} solutions. As is well known, the latter correspond
to the free oscillations of the solid ---no density of external forces in the
rhs of~\eref{eqmovKV}. As usual, we attempt to find such eigen-solutions
in the factorised form

\begin{equation}
     {\bf s}({\bf x},t)=T(t)\,{\bf s}({\bf x})    \label{separ}
\end{equation}
which results in the following (dots on symbols stand for time derivatives):

\begin{eqnarray}
  \rho\ddot T(t)\,{\bf s}({\bf x}) & = & \left[\mu T(t)+\mu'\dot T(t)\right]
  \nabla^2{\bf s}({\bf x}) \nonumber \\[1 ex]
  & + & \left[(\lambda+\mu)T(t) + (\lambda'+\mu')\dot T(t)\right]
  \nabla\left(\nabla{\bf\cdot}{\bf s}({\bf x})\right)
  \label{eqmovKV2}
\end{eqnarray}

Next we use the well known decomposition of a three dimensional vector
field into its irrotational a divergence free components~\cite{ll}:

\begin{equation}
  {\bf s}({\bf x}) = {\bf s}_l({\bf x}) + {\bf s}_t({\bf x})\ ,
  \qquad \nabla{\bf \cdot}{\bf s}_t({\bf x}) =
  \nabla\times{\bf s}_l({\bf x}) = 0
  \label{decom}
\end{equation}

By the methods described in~\ref{aA}, it can be seen that the equations
satisfied by ${\bf s}_t({\bf x})$, ${\bf s}_l({\bf x})$, and $T(t)$
are~\eref{A5a}--\eref{A6b}, i.e.,

\numparts
  \begin{eqnarray}
   &\nabla^2{\bf s}_t+{\cal K}^2\,{\bf s}_t=0 & \label{helmt.a} \\
   &\mu\,T+\mu'\,\dot T+{\cal K}^{-2}\ddot T=0 & \label{helmt.b}
  \end{eqnarray}
\endnumparts
and

\numparts
  \begin{eqnarray}
   &\nabla^2{\bf s}_l+{\cal Q}^2\,{\bf s}_l=0 & \label{helml.a} \\
   &\mu\,T+\mu'\,\dot T+{\cal Q}^{-2}\ddot T=0 & \label{helml.b}
  \end{eqnarray}
\endnumparts

Since $T(t)$ must fulfil {\it both\/} equations~\eref{helmt.b}
and~\eref{helml.b}, the {\it separation constants} ${\cal K}^2$ and
${\cal Q}^2$ are {\it not\/} independent. The binding relationship is
established after it is realised that the solution to those equations
is of the form

\begin{equation}
      T(t) = e^{\gamma t}     \label{Tt}
\end{equation}
where $\gamma\/$ is in general a {\it complex\/} quantity. We readily obtain

\numparts
  \begin{eqnarray}
   {\cal Q}^2 & \!= \! & -\rho\gamma^2\left[
   \lambda+2\mu+\gamma\,(\lambda'+2\mu')\right]^{-1}  \label{KQ.a} \\
   {\cal K}^2 & \!= \! & -\rho\gamma^2\left
   [\mu+\gamma\,\mu'\right]^{-1} \label{KQ.b}
  \end{eqnarray}
\endnumparts

The values $\gamma\/$ can possibly take on are determined by the
{\it boundary conditions}, equations~\eref{I.4}. The reader is
referred to~\ref{aB} for a detailed description of the eigenvalue
algebra of this problem. Just as in the non-dissipative case, there
are seen to be two families of eigenmodes: {\it toroidal\/} (purely
torsional) and {\it spheroidal\/}. Viscous effetcs are however small
in practice, as inferred from the narrow {\it linewidth\/} of the
measured resonances. This means that the following inequalities hold
in any cases of interest to us:

\begin{equation}
  \frac{\mu'}{\mu},\;\frac{\lambda'}{\mu}\ll\frac{1}{\omega}
  \label{approx}
\end{equation}
where $\omega\/$ is the frequency of the mode considered. We shall use these
inequalities to estimate the roots of the eigenvalue equation~\eref{eigen}
(see~\ref{aB}) {\it perturbatively\/} from the non-dissipative ones,
already known ---see \cite{lobo} for full details. Clearly thus, our
procedure will be valid for the lower frequency modes. We proceed
sequentially for the two families of eigenmodes.

\subsection{Toroidal modes}
\label{sec.3a}

These correspond to the solutions to (see~\eref{eigen})

\begin{equation}
  \beta_1({\cal K}R) = 0
  \label{torKV}
\end{equation}
which is {\it formally\/} identical to the toroidal eigenvalue equation
for a non-dissipative solid sphere~\cite{lobo}. If we call $k_{nl}^T$
the toroidal wave numbers of the latter, we have

\begin{equation}
  {\cal K}_{nl}^T = k_{nl}^T = \sqrt{\frac{\rho}{\mu}}\omega_{nl}^T
  \label{KRTKV}
\end{equation}
and hence, by equation~\eref{KQ.b},

\begin{equation}
  \gamma_{nl}^T = -(\omega_{nl}^T)^2\frac{\mu'}{2\mu} +
  i\omega_{nl}^T\sqrt{1-\left(\frac{\omega_{nl}^T\mu'}{2\mu}\right)}
  \simeq i\omega-\omega^2\frac{\mu'}{2\mu}
  \label{gammaT}
\end{equation}
where the last approximation depends on the validity of the
assumption~\eref{approx}. Expression~\eref{gammaT} nicely shows
how this Kelvin-Voigt model predicts {\it exponentially damped\/}
eigenmode oscillations. If we recall that such damping is expediently
described in terms of a {\it quality factor Q\/} ---see~\eref{I.1}
above--- then we discover that the prediction of the model is that

\begin{equation}
  Q_{nl}^T = \frac{2\mu}{\mu'}\left(\omega_{nl}^T\right)^{-1}
  \label{Qtor}
\end{equation}
i.e., the quality factor for toroidal modes is {\it inversely proportional\/}
to the frequency of the mode. The {\it amplitudes\/} of these modes have the
form

\begin{equation}
  {\bf s}^{T}_{KV}({\bf x},t)={\bf s}^{T}_{E}({\bf x},t)\,e^{-\omega t/Q}\ ,
  \qquad Q=\frac{2\mu}{\mu'\omega}
\end{equation}
where the subindex $KV\/$ stands for `Kelvin-Voigt', while $E\/$ refers
to the standard frictionless case, whose amplitudes are those given e.g.
in reference \cite{lobo}.

\subsection{Spheroidal modes}
\label{qnsphm}

A second alternative to find a non-trivial solution of the linear
system~\eref{bcatlastKV} is to impose the condition

\begin{equation}
 \beta_4\left({\cal Q} R,\frac{\lambda+\gamma\lambda'}{\mu+\gamma\mu'}\right)
 \,\beta_3({\cal K} R)-l(l+1)\,\beta_1({\cal Q} R)\beta_1({\cal K} R) = 0
 \label{sphKV}
\end{equation}

This is characteristic of the {\it spheroidal\/} eigenmodes. By virtue
of equations~\eref{KQ.a} and~\eref{KQ.b}, this relationship can be
translated into a condition to be fufilled by $\gamma$, and which
depends on the ratios $\lambda/\mu$, $\lambda'/\lambda$ and $\mu'/\mu$,
as well as on the multipole index $l\/$. In this case, as we are not
dealing with an eigenvalue problem of a selfadjoint operator, complex
solutions to equation~\eref{sphKV} are allowed ---indeed, expected.
An exact solution of that equation implies a separation of its real
and imaginary parts, followed by numerical calculations which determine
the angular frequency and quality factor of the quasinormal mode at hand.
We are interested in materials with long decay times, so we shall set up
a perturbative solution to equation~\eref{sphKV}, using 

\begin{equation}
  \epsilon\equiv\frac{\mu'\omega}{\mu}   \label{eps}
\end{equation}
as the small perturbative parameter. In other words, we assume that the
approximation~\eref{approx} holds. Here, $\omega\/$ stands for a generic
spheroidal eigenfrequency of the non-dissipative solid. Obviously, the
unperturbed solution, i.e., that corresponding to $\epsilon\/$\,=\,0, is
the elastic solid's solution, already discussed in reference \cite{lobo}.
We thus introduce the perturbative expansion

\begin{equation}
  \gamma=\gamma_0+\gamma_1\epsilon+O(\epsilon^2),\hspace{1cm}
  \gamma_0=-i\omega,\label{gamma}
\end{equation}

Using equations~\eref{KQ.a}, \eref{KQ.b} and~\eref{gamma}, we obtain
perturbative expansions for the parameters ${\cal K}$ and ${\cal Q}$
which can be written as

\begin{equation}
  {\cal K}=k_0+k_1\epsilon+O(\epsilon^2),\qquad
  {\cal Q}=q_0+q_1\epsilon+O(\epsilon^2),\label{perkq}
\end{equation}
where

\numparts
\begin{eqnarray}
  & k_0=k=\omega\sqrt{\frac{\rho}{\mu}}\ ,\qquad
    k_1=i\sqrt{\frac{\rho}{\mu}}\left(\gamma_1+\frac{\omega}{2}\right) &
    \label{kq.a} \\[1 em]
  & q_0=q=\omega\sqrt{\frac{\rho}{\lambda+2\mu}}\ ,\qquad
    q_1=i\sqrt{\frac{\rho}{\lambda+2\mu}}\left(\gamma_1+
    \frac{h'+2}{h+2}\frac{\omega}{2}\right) &
    \label{kq.b}
\end{eqnarray}
\endnumparts

In the above equations, $k$ and $q$ are the parameters appearing in the
elastic sphere's case, and we have introduced the dimensionless ratios

\begin{equation}
  h\equiv\frac{\lambda}{\mu}\ ,\qquad h'\equiv\frac{\lambda'}{\mu'}
  \label{hs}
\end{equation}
which are both zero order quantities. We can now perform the perturbative
expansion of the eigenvalue equation. In order to ease the resulting
expressions, let us introduce the following notation:

\numparts
\begin{eqnarray}
 \hspace*{-1.7 cm}
 l(l+1)\,\beta_1({\cal K} R) & = & l(l+1)\beta_1(kR)+
 l(l+1)\beta'_1(kR)k_1R\,\epsilon  \nonumber  \\
 & \equiv & B_0+B_1k_1R\,\epsilon  \\
 \hspace*{-1.7 cm}
 \beta_1({\cal Q} R) & = & \beta_1(qR)+\beta'_1(qR)q_1R\,\epsilon
 \nonumber \\
 & \equiv & C_0+C_1q_1R\,\epsilon  \\
 \hspace*{-1.7 cm}
 \beta_3({\cal K} R) & = & \beta_3(kR)+\beta'_3(kR)k_1R\,\epsilon
 \nonumber \\
 & \equiv & D_0+D_1k_1R\,\epsilon  \\
 \hspace*{-1.7 cm}
 \beta_4\left({\cal Q} R,\frac{\lambda+\gamma\lambda'}{\mu+\gamma\mu'}\right)
 & = & \beta_4(qR,h)+\left[\beta'_4(qR,\lambda/\mu)q_1R-\frac{i}{2}
 (h'-h)\,j_l(qR)\right]\epsilon  \nonumber  \\
 & \equiv & A_0+(A_1q_1R-iA'_1)\,\epsilon
 \label{bet4}
\end{eqnarray}
\endnumparts
where a prime on a $\beta\/$ function denotes differentiation with
respect to its {\it first\/} argument. We note that the functions
$A_0,\ldots,D_1$ (defined in every second line of the above set of
equations) are real. With this notation, the zero-th order form of
equation~\eref{sphKV} is

\begin{equation}
   A_0D_0-C_0B_0=0
\end{equation}
which is simply the condition that $\omega\/$ be a spheroidal eigenvalue
of the purely elastic case. On the other hand, the first order expansion
of~\eref{sphKV} yields

\begin{equation}
  (A_0D_1-C_0B_1)k_1R+(A_1D_0-C_1B_0)q_1R=iA'_1D_0,
\end{equation}
whence, using the form of $k_1$ and $q_1$, the value of $\gamma_1$ ensues.
It can be written as

\begin{equation}
  \gamma_1=-\frac{\omega}{2}\,[f(kR,h,h')]^{-1}
  \label{vgam}
\end{equation}
where we define the dimensionless function $f\/$ by\footnote{
The case in which $f\/$ takes its simplest form is that of monopole modes.
We know (see \cite{lobo}) that when $l\/$\,=\,0 equation~\eref{sphKV} is
no longer valid, and must be replaced by

\begin{equation}
  \beta_4\left({\cal Q} R,\frac{\lambda+\gamma\lambda'}{\mu+\gamma\mu'}
  \right) = 0
\end{equation}

Using now the expansion~\eref{bet4} and the fact that $A_0=0$ for the
uperturbed monopole eigenfrequencies, we obtain for the first order
correction to ${\cal Q}$,

\begin{equation}
  q_1=iA'_1A_1^{-1}
  \label{q1l0}
\end{equation}
and therefore, using the notation of equation~\eref{vgam} and the
relationships~\protect\eref{kq.a}-\protect\eref{kq.b},

\begin{equation}
  f(qR,h,h')= \left[\frac{h'+2}{h+2}-\frac{2\,A'_1}{qR\,A_1}\right]^{-1}.
\end{equation}}

\begin{eqnarray}
  \hspace*{-1.8 cm} f(kR,h,h')\equiv & & \nonumber \\
  \hspace*{-1.3 cm}
  -\frac{A_0D_1-C_0B_1+(A_1D_0-C_1B_0)(h+2)^{-1/2}}
  {2A'_1D_0(kR)^{-1}-A_0D_1+C_0B_1-(A_1D_0-C_1B_0)\,(h'+2)(h+2)^{-3/2}} & &
  \label{f}
\end{eqnarray}

We note that the first order correction obtained for $\gamma$ is {\it real}.
Therefore, to this order of approximation, the frequencies of vibration
remain unaltered, and are the same as those obtained for the elastic solid.
Moreover, $k_1$ and $q_1$ happen to be purely imaginary. Therefore, the
modulus of the radial functions appearing in the spatial part of spheroidal
quasinormal modes of vibration will also be the same as those of the
ideally elastic solid, for the corrections to $k\/$ and $q\/$ will just
introduce, to first order, a {\it complex phase\/} factor.

\begin{figure}[t]
\centering
\includegraphics[width=11cm]{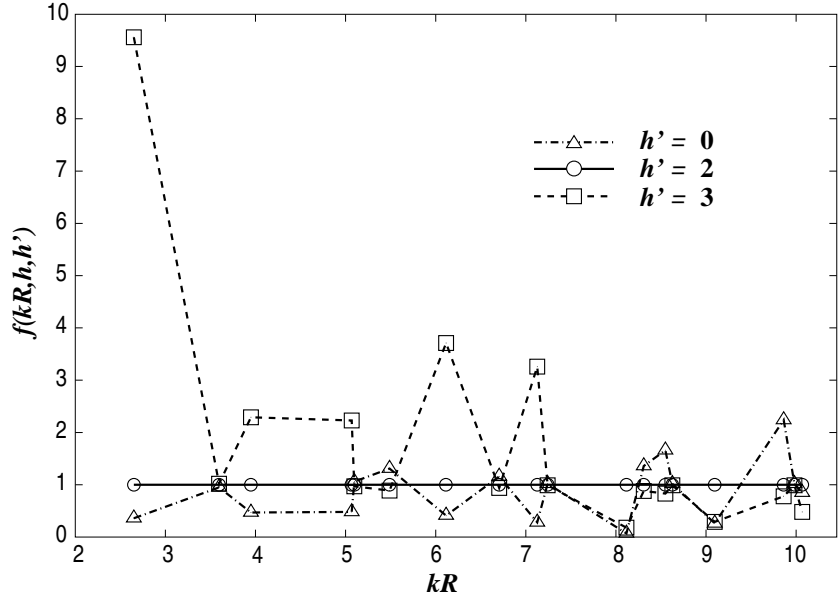}
\caption{Plot of the function $f(kR,h,h')$ ---see (\protect\ref{f})---
for a Poisson ratio $\sigma\/$\,=\,1/3 (which means $h\/$\,=\,2), and for
the first 18 (spheroidal) modes of the sphere's spectrum and a few values
of the viscoelastic ratio $h'\/$. Note that $f(kR,h,h')\equiv 1$ if $h=h'$.
\label{f1}}
\end{figure}

Summing up, we have shown that while the spheroidal normal modes of vibration
of an elastic solid are given by an expression of the form~\cite{lobo}

\begin{equation}
  {\bf s}_E^P({\bf x},t) = e^{i\omega_{nl}^Pt}\,\left[
  A_nl(r)\,Y_{lm}(\theta,\varphi)\,{\bf n}
  -B_{nl}(r)\,i{\bf n}\!\times\!{\bf L}Y_{lm}(\theta,\varphi)\right]
  \label{sEP}
\end{equation}
the spheroidal quasinormal modes ${\bf s}_{KV}^P$ of a Kelvin--Voigt solid
are obtained from the normal modes of the elastic solid according to the
following

\begin{eqnarray}
  {\bf s}_{KV}^P({\bf x},t) & = & e^{i\omega_{nl}^Pt-\omega_{nl}^Pt/Q}\,
  \left[e^{i\chi_1(r)}A_{nl}(r)Y_{lm}(\theta,\varphi)\,{\bf n}\right.
  \nonumber \\
  & - &
  \left. e^{i\chi_2(r)}B_{nl}(r)\,i{\bf n}\!\times\!
  {\bf L}Y_{lm}(\theta,\varphi)\right]
  \label{qnKV}
\end{eqnarray}
the qualitity factor being given by

\begin{equation}
  Q^P_{nl} = \frac{2\mu}{\mu'\omega_{nl}^P}\,f(k^P_{nl}R,h,h')
  \label{QQnKV}
\end{equation}
where, it is recalled, $f\/$ is given by~\eref{f} as a function
of the mode and the coefficients characterising the viscoelastic
properties of the solid. The real phases $\chi^{}_{1,2}(r)$ can
be computed from equations~\eref{sKV.1}-\eref{sKV.3}, \eref{kq.a},
and~\eref{kq.b}. Nevertheless, the explicit (and cumbersome) form of
these phases is largely irrelevant, and whe shall omit its explicit
form here~\cite{jup}. They merely introduce a position dependent
shift in the phase of the vibrations which is of order $\epsilon\/$,
therefore not likely to give rise to measurable effects. More
interesting and physically relevant is the behaviour of the function
$f\/$ giving the precise dependence of the quality factor on frequency.
First of all, it is easily seen that, for the special case
$h\/$\,=\,$h'\/$, $f\/$ is equal to~1, and thus the quality factor
is proportional to $\omega^{-1}$, as was the case with toroidal modes.
But when the aforementioned equality does not hold, numerical calculations
are needed. Figure \ref{f1} shows that this function has a rather
irregular dependence on frequency. We have represented $f\/$ for
the first 18~eigenvalues of the spheroidal spectrum, where the
inequality~\eref{approx} safely holds. It is clear from equation~\eref{f}
that holding fixed $h\/$ and $kR\/$ leaves us with a {\it linear\/}
function of $h'\/$, whose slope varies from root to root.

\section{Maxwell model}
\label{sec.4}

In this section we shall consider constituent equations given by the so
called Maxwell model. Like the Kelvin-Voigt, these equations only involve
first order time derivatives. As we shall see, quite different predictions
will be obtained for the quality factor dependence on the mode frequencies.

The Maxwell model is also isotropic and homogenous, and is characterised by
the following constituent equations \cite{ja}:

\begin{equation}
  \partial_t\sigma_{ij}
  +\alpha\,\sigma_{kk}\delta_{ij}^{}+\beta\,\sigma_{ij} =
  \frac{\partial}{\partial t}\,\left(\lambda\,s_{kk}\,\delta^{}_{ij}
  + 2\mu\,s_{ij}\right)    \label{constM}
\end{equation}

Here, the constants $\lambda$ and $\mu$ are again the Lam\'e coefficients
describing the elastic behaviour of the body, while the constants $\alpha$
and $\beta\/$ parametrise the effects due to internal friction. To construct
factorised solutions, we must also factorise both stress and strain\footnote{
Due to the equations of motion~\eref{eqmov}, if ${\bf s}({\bf x},t)$ is
assumed to be separable in the fashion~\eref{separ} then the strain
tensor is separable too, and due to the constituent equation it is
easily seen that the only possible time dependence is of the form 
\mbox{$\exp(\gamma t)$}.}:

\begin{equation}
  \sigma_{ij}({\bf x},t)=e^{\gamma t}\,\sigma_{ij}({\bf x})\ ,  \qquad
  s_i({\bf x},t) = e^{\gamma t}\, s_i({\bf x}),
\end{equation}

The constituent equation is thus written, after separation of variables
and contraction of its free indices, as

\begin{equation}
  (\gamma+3\alpha+\beta)\,\sigma_{jj}({\bf x}) =
  \gamma\,(2\mu+3\lambda)\,s_{jj}({\bf x})
\end{equation}

and hence we have the following relationship between the spatial parts of
stress and strain:

\begin{equation}
  \left(1+\frac{\beta}{\gamma}\right)\,\sigma_{ij}({\bf x}) =
  \left(\lambda-\alpha\frac{2\mu+3\lambda}{\gamma+3\alpha+\beta}\right)
  \,s_{kk}({\bf x}) \delta_{ij}^{}+2\mu\,s_{ij}({\bf x})
\end{equation}

Like in the Kelvin-Voigt model, we shall be mainly interested in the case
of small internal friction, so we shall assume

\begin{equation}
  \frac{\beta}{\mid\gamma\mid},\;\frac{\alpha}{\mid\gamma\mid}\ll 1
  \label{ApM}
\end{equation}
whence the following constituent relationship results:

\begin{equation}
  \sigma_{ij}({\bf x})=\lambda\left(1-\frac{\delta}{\gamma}\right)
  \,s_{kk}({\bf x})\delta^{}_{ij} + 2\mu\left(
  1-\frac{\beta}{\gamma}\right)\,s_{ij}({\bf x})     \label{consM}
\end{equation}
where we have introduced a new constant, $\delta$, defined by

\begin{equation}
  \delta\equiv\frac{2\mu+3\lambda}{\lambda}\,\alpha + \beta,
\end{equation}
so that we can take as the parameters characterizing the Maxwell solid
the set consisting of the Lam\'e coefficients $\lambda$, $\mu$, and the
parameters $\beta$, $\delta\/$ which describe internal friction.

Let us compare equation~\eref{consM} with that of the Kelvin--Voigt model
---cf.~\eref{consKV}--- once the separation of variables has been performed:

\begin{equation}
  \sigma_{ij}({\bf x})=(\lambda+\gamma\lambda')\,s_{kk}({\bf x})\,\delta_{ij}
  + 2(\mu+\gamma\mu')\,s_{ij}({\bf x})    \label{consKVag}
\end{equation}

Comparing equations~\eref{consM} and~\eref{consKVag}, we observe that the
solution of the Maxwell model can be carried out, as regards the spatial
part of ${\bf s}$, following the same method used in the previous section
for the Kelvin--Voigt model. In fact, we can directly take the expressions
there derived, and make the substitutions

\begin{equation}
  \mu'\ \longrightarrow\ -\mu\beta\gamma^{-2}\ ,\qquad
  \lambda'\ \longrightarrow\ -\lambda\delta\gamma^{-2}
\end{equation}
which transform equation~\eref{consKVag} into~\eref{consM}. Thus the form
of the solutions and boundary conditions for a Maxwell viscoelastic sphere
are those of~\ref{aB} with constants ${\cal K}\/$ and ${\cal Q}\/$ now
given by the following functions of the parameter $\gamma$:

\numparts
\begin{eqnarray}
  {\cal Q} & = & i\sqrt{\frac{\rho}{\lambda+2\mu}}\,\left[
  \gamma+\frac{1}{h+2}\left(\frac{h}{2}\delta+\beta\right)\right]
  \label{KQM.a}  \\[1 ex]
  {\cal K}&=&i\sqrt{\frac{\rho}{\mu}}\left(\gamma+\frac{\beta}{2}\right)
  \label{KQM.b}
\end{eqnarray}
\endnumparts
where the approximation~\eref{ApM} has been taken into account, and
$h\/$\,$\equiv$\,$\lambda/\mu$.

Thus the two families of quasinormal modes of vibration are also present
in this model, and we describe them in the following subsections.

\subsection{Toroidal modes}
\label{sec.4a}

As already discussed, the allowed values for $\gamma$ are again those
making the linear system~\eref{bcatlastKV} compatible, and there are two
alternative ways to accomplish this. The first possibility yields purely
tangential ($C_t=C_l=0$) vibrations satisfying once more the condition

\begin{equation}
  \beta_1({\cal K} R) = 0\ \ \Longrightarrow\ \ 
  {\cal K} = \sqrt{\frac{\rho}{\mu}}\,\omega_{nl}^T
\end{equation}
$\omega_{nl}^T$ being a toroidal eigenfrequency of the elastic sphere.
Using the relationship between $\gamma$ and ${\cal K}$ for a Maxwell
solid given by equations~\eref{KQM.a}-\eref{KQM.b}, we obtain the
allowed values~$\gamma_{nl}^T$ as

\begin{equation}
  \gamma_{nl}^T = -i\omega_{nl}^T-\frac{\beta}{2}    \label{gammaM}
\end{equation}

Again, toroidal quasinormal modes have two fundamental properties: they
have the same set of eigenfrequencies as the elastic sphere (to first order
in the parameters describing internal friction, $\beta$ in this case), and
also exactly the same spatial part (for {\em all\/} values of the viscosity
parameters). The only difference between Kelvin--Voigt and Maxwell solids as
regards toroidal modes appears in the dependence of the quality factor on
$\omega$: as equation~\eref{gammaM} shows, the quality factor in a linear
Maxwell solid increases linearly with frequency. We can express all these
properties by means of the following formul\ae:

\begin{equation}
  {\bf s}_{M}^T({\bf x},t) = {\bf s}_{E}^T({\bf x},t)
  \,e^{-\omega_{nl}^Tt/Q}\ ,\qquad
  Q^T_{nl} = \frac{2\omega_{nl}^T}{\beta}
  \label{4.13}
\end{equation}
relating Maxwell quasi-normal modes of vibration, ${\bf s}_M^T({\bf x},t)$,
to elastic normal modes, ${\bf s}_E^T({\bf x},t)$, for the toroidal family.

\subsection{Spheroidal modes}
\label{sec.4b}

In order to handle the spheroidal family, we shall resort again to the
perturbative expansions already used in the Kelvin--Voigt case, and also
in the toroidal family, just described. The Maxwell model trivially reduces
to the perfect elastic case when $\beta\/$\,=\,$\delta$\,=\,0, hence we can
take as the perturbative parameter

\begin{equation}
  \epsilon = \frac{\beta}{\omega}
\end{equation}
where $\omega\/$ is the elastic eigenfrequency to which $\gamma\/$ approaches
when both $\beta\/$ and $\delta\/$ approach zero. Perturbative expansions in
the fashion of section~\ref{qnsphm} can now be introduced:

\begin{equation}
  \gamma = -i\omega+\gamma_1\epsilon\ ,\qquad
  {\cal K} = k+k_1\epsilon\ ,\qquad
  {\cal Q}=q+q_1\epsilon
\end{equation}
where the first order corrections $k_1$ and $q_1$ are given by
equations~\eref{KQM.a}-\eref{KQM.b} as functions of~$\gamma_1$:

\numparts
 \begin{eqnarray}
  k_1 & = & i\sqrt{\frac{\rho}{\mu}}\left(\gamma_1+\frac{\omega}{2}\right)
   \label{kqM.a}  \\[1 ex]
  q_1 & = & i\sqrt{\frac{\rho}{\lambda+2\mu}}\left(\gamma_1 +
          \frac{h'+2}{h+2}\frac{\omega}{2}\right)
   \label{kqM.b}
 \end{eqnarray}
\endnumparts

The zero--order ratio $h'\/$ is now given by

\begin{equation}
  h' = \frac{\alpha}{\beta}\,h
\end{equation}

With this definition, together with that of the perturbative parameter,
the expressions at hand are formally identical to those of the Kelvin--Voigt
model, and therefore the solutions to the Maxwell model share all their
properties with their Kelvin--Voigt counterparts; the exception is the
dependence of the quality factor on frequency: the product $\gamma_1\epsilon$,
which gives the exponential decay, is now {\it independent~of\/}~$\omega$. 

Summing up, spheroidal quasinormal modes of the Maxwell solid,
${\bf s}_{M}^P({\bf x},t)$, are related to spheroidal normal modes of a
perfectly elastic sphere by the equations

\begin{equation}
  \hspace*{-1 cm}
  {\bf s}_{M}^P({\bf x},t) = {\bf s}_{E}^P({\bf x},t)
  \,e^{-\omega_{nl}^Pt/Q+\chi(r)}\ ,\qquad
  Q^P_{nl} = \frac{2\omega_{nl}^P}{\beta}\,f(k^P_{nl}R,h,h')
  \label{qnmM}
\end{equation}
where the function $f(kR,h,h')$ is again given by~\eref{f}.

As we see, the only difference between the behaviour of Maxwell and
Kelvin--Voigt viscoelastic solids, when the internal friction effects can
be considered small, appears in the dependence of $Q\/$ on frequency. We
must however stress that, under other conditions (e.g. static load), both
models show larger divergences in their physical properties~\cite{GI}.

\section{Other Models}
\label{sec.5}

In this section we review other models which have been proposed to address
the dynamics of a viscoelastic solid. They are generalisations of those in
the two previous sections. We shall however not attempt to find complete
solutions to all of them, as it eventually becomes too cumbersome. We shall
however discuss in this section some of their most relevant traits.

\subsection{The Standard Linear Model}
\label{sec.5a}

The Standard Linear Model (SLM) for a viscoelastic solid is a generalised
combination of the Kelvin--Voigt and Maxwell models. The constituent
equations take here the form:

\begin{equation}
  \hspace*{-1.4 cm}
  \sigma_{ij} + \frac{\partial}{\partial t}\,(\alpha\,\sigma_{kk}\delta_{ij}
  + 2\beta\,\sigma_{ij}) = \left(\lambda+\alpha'\frac{\partial}{\partial t}
  \right)s_{kk}\delta_{ij} + 2\left(\mu+\beta'\frac{\partial}{\partial t}
  \right)s_{ij}    \label{consSLM}
\end{equation}
where the effects of internal friction are described in this case with
the aid of four constant parameters: $\alpha$, $\beta$, $\alpha'$, and
$\beta'$. When looking for factorised solutions, the equations of motion
and the above relationship force a time dependence of the form $e^{\gamma t}$
for both stress and strain. When such a dependence is introduced in
equation~\eref{consSLM}, we obtain the following relationship between
the spatial parts of the stress and strain tensors:

\begin{eqnarray}
  \hspace*{-1.4 cm}
  (1+2\gamma\beta)\,\sigma_{ij}({\bf x}) & = & \left[\lambda
  + \alpha'\gamma - \alpha\gamma\,
  \frac{3\lambda+2\mu+\gamma(3\alpha'+2\beta)'}{1+\gamma
  (3\alpha+2\beta)}\right]s_{kk}({\bf x})\,\delta_{ij} \nonumber \\[1 ex]
  & + & 2(\mu+\beta'\gamma)\,s_{ij}({\bf x})
\end{eqnarray}

The case of small internal friction is treated by first order approximation
in the quantities parametrising viscous processes, i.e.,

\begin{equation}
  \alpha,\;\beta,\;\alpha',\;\beta'\;\ll\;\mid\gamma\mid^{-1}
  \label{little}
\end{equation}

When such approximation is made, the above equation reduces to

\begin{equation}
  \sigma_{ij}({\bf x}) = (\lambda+\lambda'\gamma)s_{kk}({\bf x})\delta_{ij}
  + 2(\mu+\mu'\gamma)s_{ij}({\bf x})
\end{equation}
where we have introduced two new constants given by

\begin{equation}
  \mu'\equiv\beta'-2\beta\mu\ ,\qquad
  \lambda'\equiv\alpha'-2\lambda\beta - (3\lambda+2\mu)\alpha
  \label{lm}
\end{equation}

Therefore when equation~\eref{little} holds, the SLM reduces to a
Kelvin--Voigt model, i.e., for small internal friction, both models have
the same set of quasinormal modes of vibration, which are characterized
by the constants $\lambda$, $\mu$, $\lambda'$ and $\mu'$, the latter being
given, for the Standard Linear solid, by equations~\eref{lm}.

\subsection{Generalised mechanical models}
\label{sec.5b}

The models analyzed so far are the simplest ones obtained by three dimensional
generalisations of mechanical viscoelastic models composed of linear springs.
They give rise to differential constituent relations, with time derivatives
up to the first order. Considering more involved generalisations yields
differential relations involving higher order time derivatives of strain and
stress ---see e.g. \cite{HA}. Thus, quite independently of any reference to
the underlying mechanical model, we can consider general differential
relations between stress and strain including any number of time derivatives.
To ease the formulation of such differential constituent equations for the
case of isotropic and homogenous solids, we shall introduce the trace-free
parts of the strain and stress tensors, $s'_{ij}$, $\sigma'_{ij}$ (usually
termed {\em deviatoric\/} components in the literature on viscoelasticy
\cite{HA}), and their traces, $s\/$ and $\sigma\/$ ({\em dilational\/}
components), defined by

\numparts
\begin{eqnarray}
  s'_{ij} & = & s_{ij} - \frac{1}{3}\,s\,\delta_{ij}\ ,\qquad
  s \equiv s_{kk} \\[1 ex]
  \sigma'_{ij} & = & s_{ij} - \frac{1}{3}\sigma\delta_{ij}\ ,\qquad
  \sigma \equiv \sigma_{kk}.
\end{eqnarray}
\endnumparts

In terms of the above quantities, the linear Hooke law for an elastic solid
takes the form

\begin{equation}
  \sigma = (3\lambda+2\mu)s\hspace{1cm}\sigma'_{ij} =
  2\mu\,s'_{ij}
  \label{conselast}
\end{equation}
while the constituent equation of an SLM is written

\numparts
\begin{eqnarray}
  & \left[1+(3\alpha+2\beta)\,\frac{\partial}{\partial t}\right]\,\sigma =
  \left[(3\lambda+2\mu) + (3\alpha'+2\beta')\,
  \frac{\partial}{\partial t}\right]\,s & \\[1 ex]
  & \left(1+2\beta\,\frac{\partial}{\partial t}\right)\,\sigma'_{ij} =
  \left(2\mu+2\beta'\,\frac{\partial}{\partial t}\right)\,s'_{ij} &
\end{eqnarray}
\endnumparts

This equation can now be generalised to include higher order time
derivatives. We can thus consider viscoelastic models whose constituent
equation is given by

\begin{equation}
  R(\partial_t)\sigma = S(\partial_t)s\ ,\qquad
  R'(\partial_t)\sigma'_{ij} = S'(\partial_t)s'_{ij}   \label{consgen}
\end{equation}
where we have introduced formal polynomials

\numparts
\begin{eqnarray}
   & R(x) \equiv\sum_{l=0}^{N-1}r_l x^l\ ,\quad
   R'(x)\equiv\sum_{l=0}^{N-1}r'_lx^l &   \label{poly.a} \\
   & S(x) \equiv\sum_{l=0}^{N-1}s_l x^l\ ,\quad
   S'(x)\equiv\sum_{l=0}^{N-1}s'_lx^l &
   \label{poly.b}
\end{eqnarray}
\endnumparts
so that a general differential model is given for each set of 4$N\/$ real
constants $r_l\/$, $r'_l\/$, $s_l\/$, and $s'_l\/$ characterising the solid.
Some of these constants may vanish. The Kelvin-Voigt, Maxwell and SL models
considered above are of course special cases within this general class.
Several procedures have been proposed in the literature to solve the general
equations ---see \cite{MA} for a review and further reference---, which can
be applied to the solid viscoelastic sphere problem. We now sketch how they
work in this case of our interest.

Let $\tilde\sigma_{ij}({\bf x},\Omega)$ and $\tilde s_{ij}({\bf x},\Omega)$
be the Fourier transforms of the stress and strain tensors, and
$\tilde s_i({\bf x},\Omega)$ that of the displacement vector field:

\numparts
 \begin{eqnarray}
  \tilde\sigma_{ij}({\bf x},\Omega) & \equiv &
     \int_{-\infty}^\infty\sigma_{ij}({\bf x},t)\,e^{i\Omega t}\,dt
     \label{ft.a}  \\
  \tilde s_i({\bf x},\Omega) & \equiv &
     \int_{-\infty}^\infty s_{i}({\bf x},t)\,e^{i\Omega t}\,dt
     \label{ft.b}  \\
  \tilde s_{ij}({\bf x},\Omega) & \equiv &
     \int_{-\infty}^{\infty}s_{ij}({\bf x},t)\,e^{i\Omega t}\,dt
     \label{ft.c}
 \end{eqnarray}
\endnumparts

In terms of these, Fourier transforms the constituent equations
\eref{consgen} read:

\begin{equation}
  \tilde\sigma = \frac{R(i\Omega)}{S(i\Omega)}\,\tilde s\ ,\qquad
  \tilde\sigma'_{ij} = \frac{R'(i\Omega)}{S'(i\Omega)}\,\tilde s_{ij}
  \label{constran}
\end{equation}
while the equations of motion are 

\begin{equation}
  -\Omega^2\tilde{\bf s} = \frac{1}{3}\left(\frac{R(i\Omega)}{S(i\Omega)} -
   \frac{R'(i\Omega)}{S'(i\Omega)}\right)\nabla(\nabla{\bf \cdot}
   \tilde{\bf s}) + \frac{R'(i\Omega)}{2S'(i\Omega)}\,\nabla^2\tilde{\bf s}
\end{equation}

Comparing the above equations with the corresponding ones for normal modes
of vibration of elastic solids, and the constituent relation~\eref{constran}
with~\ref{conselast}, we note that the problem of finding solutions to the
equation of motion of a general viscoelastic differential model reduces to
that of finding the normal modes of vibration of an elastic solid having
{\it complex\/} Lam\'e coefficients given by

\begin{equation}
  \tilde\lambda(\Omega) = \frac{1}{3}\,\left(\frac{R(i\Omega)}{S(i\Omega)}
  - \frac{R'(i\Omega)}{S'(i\Omega)}\right)\ ,\qquad
  \tilde\mu(\Omega) = \frac{1}{2}\frac{R'(i\Omega)}{S'(i\Omega)}
\end{equation}
where the allowed values of $\Omega$ are obtained as the solutions
to the elastic solid's eigenfrequency equation when the above complex
coefficients are used instead of the real, constant Lam\'e coefficients
$\lambda$, $\mu$. Generally, $\Omega$ will have complex values, thus
giving rise to damped system oscillations. After solving for $\Omega$,
the spatial part of the solutions is obtained from that of the normal
modes by simply substituting the old, real--valued constants $\omega$,
$\lambda$ and $\mu$ by the new complex values $\Omega$, $\tilde\lambda$
and $\tilde\mu$. This method for solving the viscoelasticity is often
termed in the literature on the subject the {\em Correspondence
Principle}~\cite{MA}, and as a matter of fact our previous derivations
of the form of the quasinormal modes for Kelvin-Voigt, Maxwell and SL
models can be seen to be special cases of its application. The method
is applicable to any boundary value problem whose elastic counterpart
is solvable. The case of small internal friction (i.e., first order
approximation in the coefficients of the polynomials
\eref{poly.a}-\eref{poly.b} has been considered by Graffi~\cite{MA}
for one dimensional wave propagation.

The three dimensional spherical case is also solvable, as we know. The
{\it toroidal\/} modes are relatively straightforward to obtain from their
elastic counterparts due to the simple form of their eigenvalue equation,
while the spheroidal ones demand more complex algebra, which becomes
increasingly cumbersome as the order $N\/$ of the model increases.
We shall present here the general solution for the toroidal modes for any
differential viscoelastic model, whereby we shall obtain the dependence of
their $Q\/$ on frequency. This will also be the approximate dependence for
the spheroidal modes, if friction effects are small, as was the case with
the first order models analysed so far. A complete solution for the latter
modes can also be systematically found, but will be omitted due to its
scarcely useful algebraic complexity \cite{jup}.

\subsubsection{Toroidal modes}
\label{sec.5c}

As discussed above, the boundary equation for the toroidal modes in a
general viscoelastic model is obtained from the eigenvalue equation of the
elastic model:

\begin{equation}
  \beta_1(kR) = 0 \ ,\qquad   k = \sqrt{\frac{\rho}{\mu}}\,\omega
  \label{alteigen}
\end{equation}

Upon substitution of $\mu$ by $\tilde\mu$, we obtain

\begin{equation}
  \beta_1({\cal K}R) = 0\ ,\qquad
  {\cal K} = \sqrt{\frac{2\rho S'(i\Omega)}{R'(i\Omega)}}\,\Omega
\end{equation}

We know that the only solutions to the eigenvalue equation~\ref{alteigen}
are the {\it real\/} eigenfrequencies of the elastic sphere $\omega_{nl}^T$,
and therefore the allowed values for $\Omega$ are given by the implicit
relationship

\begin{equation}
  \sqrt{\frac{2\mu\,S'(i\Omega)}{R'(i\Omega)}}\,\Omega = \omega_{nl}^T
  \label{jqs}
\end{equation}

Let us now write the polynomials $S'\/$ and $R'\/$ in the form

\begin{equation}
  S'(x) = 1 + \epsilon\sum_{l=1}^{N-1} s'_lx^l\ ,\qquad
  R'(x) = 2\mu\left(1+\epsilon\sum_{l=1}^{N-1} r'_l {x}^l\right)
\end{equation}
as a suitable one to imply that internal friction effects are small, letting

\begin{equation}
   \epsilon \ll 1
\end{equation}

The quantities $r'_l\omega^l\/$ and $s'_l\omega^l\/$ are thus zero order
in $\epsilon$ and dimensionless, $\omega$ being a toroidal eigenvalue of
the elastic case. We then introduce an expansion for $\Omega$ in the
small parameter $\epsilon$, whose zeroth order term corresponds to a given
toroidal eigenfrequency $\omega\/$ of the elastic solid:

\begin{equation}
  \Omega = \omega + \epsilon\Omega_1
\end{equation}

Under the above conditions, we have

\begin{equation}
  \frac{2S'(i\Omega)}{\mu R'(i\Omega)} = 1 + 2i\epsilon\sum_{l=0}^{N-1}
   t_l\omega^l\ ,\qquad  t_l = i^{l-1}(s'_l-r'_l)/2
\end{equation}
and the value of $\Omega_1$ follows when introducing the above expansion
into equation~\eref{jqs}, yielding

\begin{equation}
   \Omega_1 = \sum_{l=1}^{N-1} t_l\omega^{l+1}
\end{equation}

In terms of $\Omega_1$, the quality factor reads

\begin{equation}
  Q = -\frac{\omega}{\epsilon}\,\frac{1}{\mbox{Im}[\Omega_1]}
\end{equation}

where Im[$\cdot$] denotes the imaginary part of its argument. Thus we
observe that, as regards toroidal modes, using a general differential
model gives us a {\it polynomial\/} in $\omega\/$ for $1/Q\/$, with no
independent term, so that constant $Q\/$ is not allowed by these models.
The polynomial only contains {\it odd\/} powers of the unperturbed
frequency $\omega$. In general, whenever $t_l\/$\,$\neq$\,0 for even $l\/$,
the real part of $\Omega_1$ will not vanish, and the angular frequency of
the periodic component of the quasinormal modes shall undergo first order
corrections. Hence, in order to preserve the elastic spectrum to first
order, our model must satisfy the conditions

\begin{equation}
   t_l = 0 \qquad \mbox{($l\/$ even)}
\end{equation}

Provided the preceeding equation holds, the corrections to ${\cal K}$
will be purely imaginary, and therfore the modulus of the spatial part
of the modes will remain unaltered, the only effect of viscosity being
the addition of a point dependent phase in the fashion of
equation~\eref{qnmM}\footnote{
This correction will only appear provided that $N\/$\,$\geq$\,3. This is
why it was absent in the toroidal families of the previously discussed
models.}.

The calculation for the spheroidal quasinormal modes can be performed along
the same lines but, as we have seen, the algebra is considerably more
involved already in the simplest models. It does naturally become more
cumbersome as the order of the model increases, so we omit a detailed
discussion of its technicalities here.

\section{Experimental data}
\label{sec.6}

We now present a confrontation of the above theoretical models with
available experimental measurements, some of them published~\cite{cffth},
and others unpublished as of this date~\cite{arl}.

It is very important, in order to rightly assess the suitability of
the theoretical models, to keep in mind that measurements of $Q\/$
are intrinsically difficult. The most important source of problems
is the fact that almost any experimental environment conditions
(suspensions, readout, electronics, etc.) significantly couple to
the primary oscillator (the sphere itself), and almost invariably
result in a \emph{degradation} of the measured $Q$. The way and
intensity in which this happens is however utterly unpredictable.
As a rule, theoretical $Q\/$ values should thus be expected
\emph{larger} than measured ones.

\begin{figure}[t]
\centering
\includegraphics[width=9.5cm]{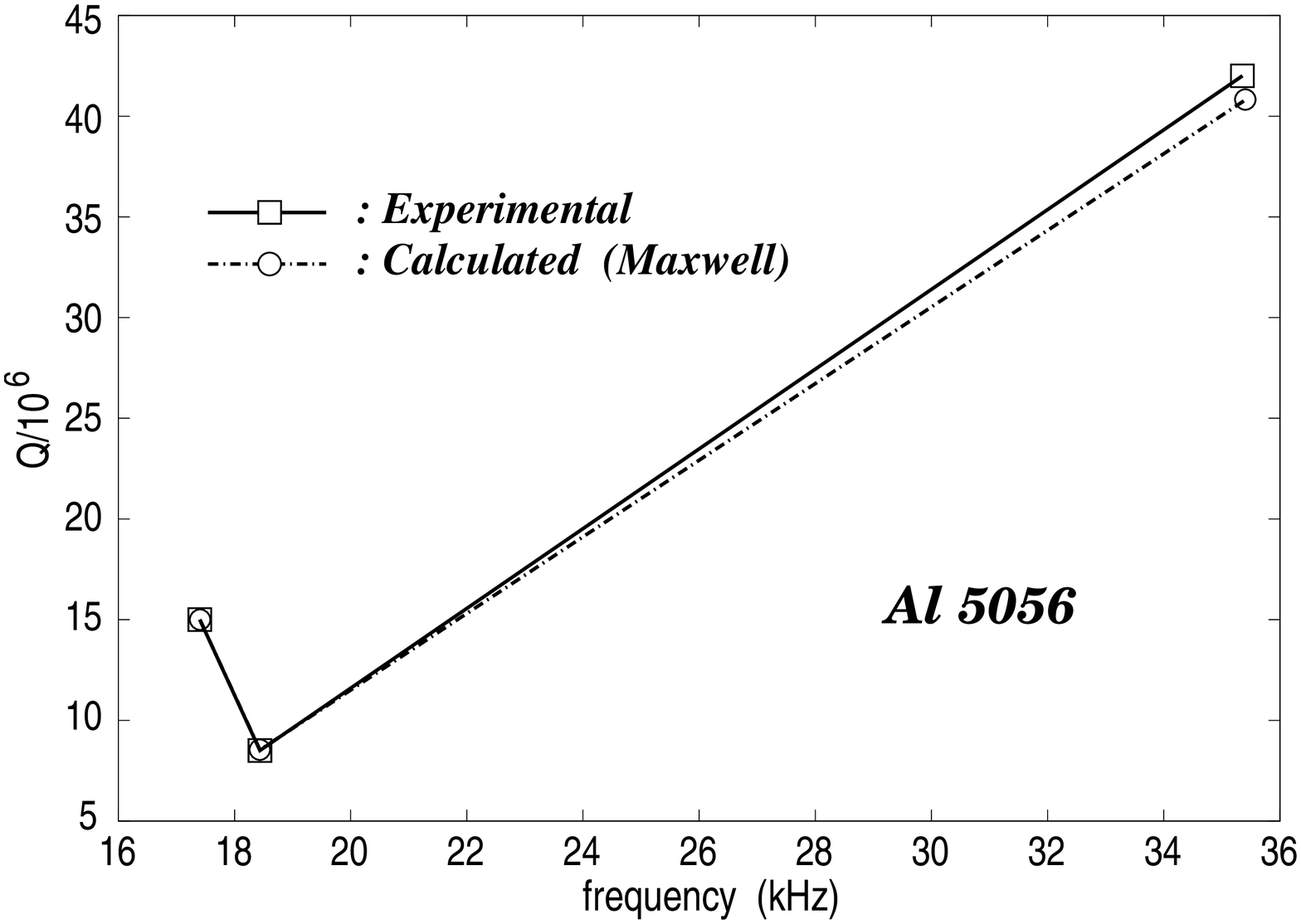}
\includegraphics[width=9.5cm]{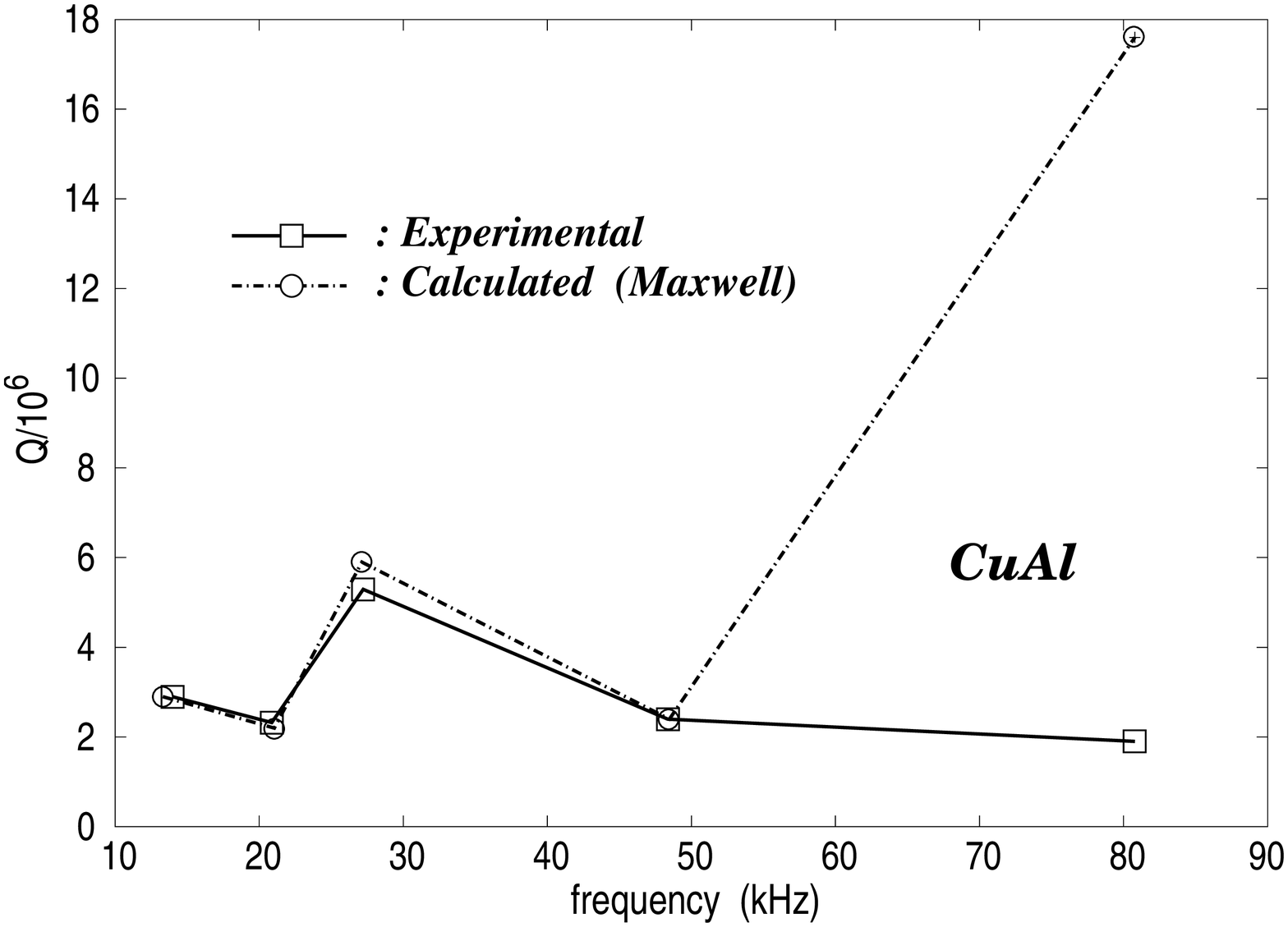}
\caption{Quality factors of a sample sphere of Al\,5056 (top), and a 7\,\%
alloy of CuAl (bottom). The first had a diameter of 153 mm, and measurements
were taken at 80 mK, while the second was 150 mm in diameter and had a
temperature of 20 mK. As can be appreciated, the $Q\/$'s of the lower
frequency modes match the theoretical predictions fairly well, assuming
a Maxwell model with viscoelastic ratios $h'=1$ and $h'=0.08$, respectively.
The higher mode in CuAl, however, considerably deviates from calculations.
\label{f2}}
\end{figure}

\begin{figure}[t]
\centering
\includegraphics[width=11cm]{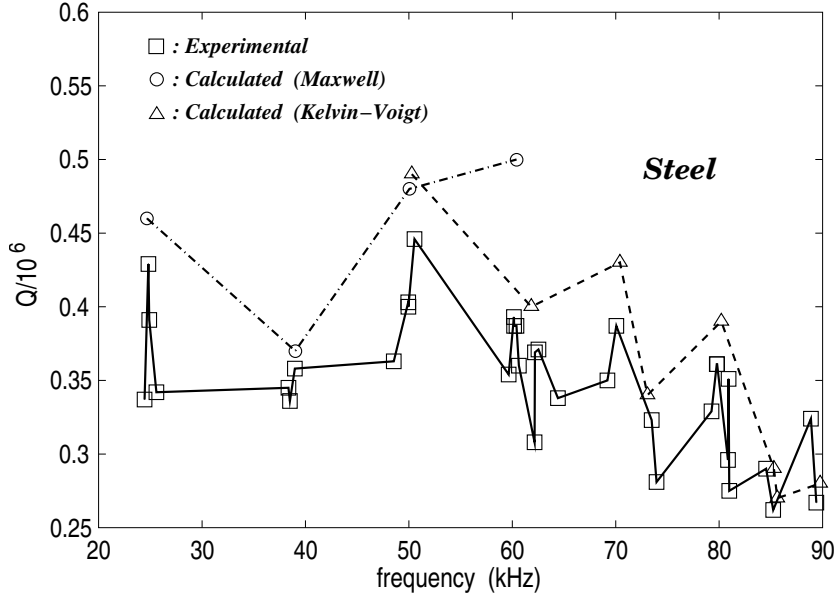}
\caption{Quality factors of a sample sphere of steel; the sample had
a diameter of 107 mm and measurements were taken at a tempreature of
4 K. Lower frequency $Q\/$'s are fairly well fit by a Maxwell model
with a viscoelastic ratio $h'=0.3$, while higher frequency $Q\/$'s
require a Kelvin--Voigt with viscoelastic ratio $h'=2.5$. In this
case there more experimental points than theoretical due to mode
splitting caused by the sphere suspension ---see text.
\label{f3}}
\end{figure}

Then, the mode frequencies slightly split up as an unavoidable
consequence of the necessity to \emph{suspend} the sample in the
laboratory ---a fact not considered in the above models. This,
at times, generates a forest of frequency peaks when different
multipole modes happen to be close to one another. For example,
within a mere 0.8\,\% of their nominal value, we find the three
fundamental frequencies: $\omega^P_{22}$, $\omega^P_{14}$, and
$\omega^T_{14}$ which, if split up, result in as many as 23~Fourier
peaks in a very reduced frequency interval. On the other hand, however,
the $Q\/$'s of the various multiplet components often group around
clearly distinct values, close to those predicted by theory. In such
cases, therefore, the viscoelastic model is helpful to tell the different
multiplet members from one another, an otherwise very difficult task.

Rather than characterising a given viscoelastic model by the two
parameters $\lambda'$ and $\mu'$, or $\alpha$ and $\beta\/$ ---see
sections~\ref{sec.3} and~\ref{sec.4}---, it is expedient for the
purposes of this section to use a different parameter pair. This
will be chosen as the $Q\/$ of the lowest frequency toroidal mode,
$Q_0\equiv Q^T_{12}$, and the viscoelastic ratio $h'$ of
sections~\ref{qnsphm} and~\ref{sec.4b}. The following expressions
thus hold:

\numparts
 \begin{eqnarray}
  \mbox{\sf Kelvin--Voigt:} & \quad &
  \left\{\begin{array}{ll}
  Q^T_{nl} = \left(\lf{\omega^T_{12}}{\omega^T_{nl}}\right)\,Q_0 \\[1 em]
  Q^P_{nl} = \left(\lf{\omega^T_{12}}{\omega^P_{nl}}\right)\,
		f(k^P_{nl}R,h,h')\,Q_0
  \end{array}\right.  \label{equiq.a} \\[1.2 em]
  \mbox{\sf Maxwell:} & \quad &
  \left\{\begin{array}{ll}
  Q^T_{nl} = \left(\lf{\omega^T_{nl}}{\omega^T_{12}}\right)\,Q_0 \\[1 em]
  Q^P_{nl} = \left(\lf{\omega^P_{nl}}{\omega^T_{12}}\right)\,
		f(k^P_{nl}R,h,h')\,Q_0
  \end{array}\right. \label{equiq.b}
 \end{eqnarray}
\endnumparts
obviously a rewrite of the corresponding ones in the respective sections.

In figure~\ref{f2} we plot the measured versus theoretically calculated
values of $Q\/$ for two sample spheres of Al\,5056 and a 7\,\% alloy of
CuAl ---see more technical data in the caption to the figure. Agreement
is quite good, when a Maxwell model is used, except for the higher
frequency mode, reported in the case of CuAl. It appears that a
better fit can be accomplished when a combination of both Maxwell
and Kelvin--Voigt models is used, such as we see in figure~\ref{f3},
where $Q\/$'s of a sample sphere of steel are considered. The latter
plot contains more experimental points than theoretical, and we see
here a clear example of frequency splittings due to symmetry breaking
caused by suspension, normally a diametral or semi-diametral bore
drilled across the sphere~\cite{cffth}.

\section{Conclusions}
\label{sec.7}

In this paper we have addressed the problem of whether it is possible to
systematically characterise the linewidths or, equivalently, the $Q\/$'s
of the oscillation eigenmodes of a given spherical GW detector. To this
end we have considered various phenomenological models, selected from
the specialised literature on the subject, and solved the equations of
motion in the cases of our interest. Different models are seen to predict
different  frequency dependences of the quality factors for the lower
modes, which are the ones we have paid attention to, and the ones
relevant for GW detection purposes. For example, in a Kelvin--Voigt
solid the $Q\/$ of a given mode appears to be inversely proportional
to its frequency, while in a Maxwell solid it is directly proportional
to it.

As we have seen, however, the behaviour of a given material is generally
not well described by a \emph{single} such model but rather by a suitable
combination of e.g.\ Maxwell and Kelvin--Voigt models, depending on the
range of frequencies considered. This is not too surprising, as these
models are \emph{phenomenological}, i.e., they are not based on a detailed
analysis of the \emph{physics} of the various dissipation mechanisms which
are responsible for the loss processes~\cite{bramp}, but on more or less
plausible (though not unique) generalisations of the ``friction is
proportional to velocity'' principle, which so accurately holds in
a simple unidimensional harmonic oscillator.

In spite of these limitations, and also in spite of the experimental
difficulties inherent in the determination of $Q$, it appears that
our treatment of the problem gives results which can be considered
quite good, in view of the above limitations. A significant bonus
of the present analysis is the possibility it offers to identify
nearby frequencies in the sphere spectrum by their distinct value
of $Q$, as predicted by theory. This is useful in GW detection
science since, for example, the \emph{second} spheroidal quadrupole
frequency, $\omega^P_{22}$, is very important due to large cross
section for GW energy absorption in that mode~\cite{clo} yet it is
very close (in frequency) to other modes with different $Q\/$'s.

Improvements on the results presented in this paper are certainly
possible. But we feel they would have to be based on a different
methodology, with more emphasis \emph{ab initio} on Solid State
lore, and/or on Materials Science.

\ack{
We express our thanks to J.A.\ Ortega, who contributed most of the
analytic work reported in this paper. Three of us (JAL, GF and AdW)
acknowledg hospitality in the {\sl INFN\/} Frascati Laboratories,
and support from the {\sl TARI\/} programme, contract number
{\tt HPRI-CT-1999-00088}.}

\appendix

\section{}\label{aA}

A well-behaved three-dimensional vector field ${\bf s}({\bf x})$ can be
expressed as the sum of an irrotational, ${\bf s}_l({\bf x})$, and a
divergence free, ${\bf s}_t({\bf x})$, vector fields, respectively
called the {\it longitudinal\/} and {\it transverse\/} components of
${\bf s}({\bf x})$ \cite{ll}:

\begin{equation}
  {\bf s}({\bf x}) = {\bf s}_l({\bf x}) + {\bf s}_t({\bf x})\ ,
  \qquad \nabla{\bf \cdot}{\bf s}_t=\nabla\times{\bf s}_l = 0  \label{A1}
\end{equation}

We now replace this decomposition into equation~\eref{eqmovKV2} to find

\begin{eqnarray}
  \rho\ddot T(t) ({\bf s}_t+{\bf s}_l) & = &
  \left[\mu T(t)+\mu'\dot T(t)\right]\,
  \nabla^2({\bf s}_t+{\bf s}_l)  \nonumber \\[1 ex]
  & + & \left[(\lambda+\mu)\,T(t)+(\lambda'+\mu')\,\dot T(t)\right]
  \nabla\left(\nabla\!\cdot\!{\bf s}_l\right)
  \label{A2}
\end{eqnarray}

Taking the rotational of this equation,

\begin{equation}
  \nabla\times\left[\rho\ddot T\,{\bf s}_t-(\mu\,T+\mu'\,\dot T)
  \,\nabla^2{\bf s}_t\right] = 0
  \label{A3}
\end{equation} 

The vector between square brackets is thus divergence--free and irrotational,
so it vanishes. We have therefore:

\begin{equation}
  \nabla^2{\bf s}_t=\left\{\frac{\rho\,\ddot T}{\mu\,T+\mu'\,\dot T}\right\}
  \,{\bf s}_t   \label{A4}
\end{equation}

Since the left hand side of the above equation does not depend on time, the
term between braces in the right hand side must equal a (complex) constant,
say $-{\cal K}^2$. Thus,

  \begin{eqnarray}
    & \nabla^2{\bf s}_t+{\cal K}^2\,{\bf s}_t = 0 &  \label{A5a} \\
    & \mu\,T+\mu'\,\dot T+{\cal K}^{-2}\ddot T=0 &  \label{A5b}
  \end{eqnarray}

An analogous procedure, after taking the divergence of equation~\eref{A2},
gives us the corresponding formul\ae\/ for the longitudinal part:

\begin{eqnarray}
  & \nabla^2{\bf s}_l+{\cal Q}^2\,{\bf s}_l = 0 &  \label{A6a} \\
  & (\lambda+2\mu)\,T+(\lambda'+2\mu')\,\dot T+{\cal Q}^{-2}\ddot T = 0 &
  \label{A6b}
\end{eqnarray}

where ${\cal Q}^2$ stands for another complex separation constant.

\section{}\label{aB}

We describe in this appendix the algebraic operations which lead to
the solution to the eigenvalue problem in a viscoelastic sphere.
Equations~\eref{eqmov} ought to be solved, subject to the boundary
conditions~\eref{I.4}. The latter can be cast in explicit vector form:

\begin{equation}
  \hspace*{-1.9 cm}
  (\lambda+\gamma\lambda')\left[\nabla\!\cdot\!{\bf s}({\bf x})\right]
  {\bf n} + 2(\mu+\gamma\mu')\,({\bf n}\!\cdot\!\nabla){\bf s}({\bf x}) +
  2(\mu+\gamma\mu')\,{\bf n}\!\times\!\left[\nabla\!\times\!{\bf s}({\bf x})
  \right] = 0
  \label{BCondKV}
\end{equation}

The irrotational and divergence free components, ${\bf s}_t({\bf x})$ and
${\bf s}_l({\bf x})$, can be expressed by means of auxiliary functions
$\phi({\bf x})$ and $\psi({\bf x})$:

\begin{eqnarray}
  {\bf s}_l({\bf x}) & = & {\cal Q}^{-1}C_0\,\nabla\,\phi({\bf x})
  \label{totKV.a} \\
  {\bf s}_t({\bf x}) & = & i{\cal K}^{-1}C_1\,
  \nabla\!\times\!{\bf L}\psi({\bf x}) + iC_2\,{\bf L}\psi({\bf x})
  \label{totKV.b}
\end{eqnarray}
where $C_0$, $C_1\/$ and $C_2\/$ are (so far) undetermined integration
constants, and {\bf L}\,$\equiv$\,$-i${\bf x}$\times$$\nabla$ is
the ``angular momentum'' operator. Upon substitution
of~\eref{totKV.a}-\eref{totKV.b} into~\eref{eqmov} it is readily
seen that the functions $\phi\/$ and $\psi\/$ are themselves also
solutions to corresponding Helmholtz equations,

\begin{equation}
  \nabla^2\phi({\bf x})+{\cal Q}^2\phi({\bf x}) = 0\ , \ \quad
  {\rm and}\ \quad
  \nabla^2\psi({\bf x})+{\cal K}^2\psi({\bf x}) = 0
  \label{pf}
\end{equation}

They have therefore the general form, using spherical coordinates
($r$,$\theta$,$\varphi$) for the vector {\bf x},

\begin{equation}
  \phi({\bf x}) = j_l({\cal Q}r)\,Y_{lm}(\theta,\varphi)\ ,
  \quad\ {\rm and}\ \quad
  \psi({\bf x}) = j_l({\cal K}r)\,Y_{lm}(\theta,\varphi)
  \label{pfsol}
\end{equation}

where $j_l\/$ are {\it spherical\/} Bessel functions of the first kind and
$Y_{lm}\/$ are spherical harmonics. The solutions~\eref{pfsol} are those
possessing regularity properties in the whole interior and boundary of the
solid. We thus have

  \begin{eqnarray}
  \nabla\phi & \ = & \frac{d\,j_l({\cal Q}r)}{dr}\,Y_{lm}(\theta,\varphi)\,
  {\bf n} - \frac{j_l({\cal Q}r)}{r}
  \,i{\bf n}\times{\bf L}Y_{lm}(\theta,\varphi)
  \label{sKV.1} \\
  \nabla\!\times\!{\bf L}\psi & \ = &
  -l(l+1)\,\frac{j_l({\cal K}r)}{r}\,Y_{lm}(\theta,\varphi)\,{\bf n}
  \nonumber \\
  & \ + &
  \left[\frac{j_l({\cal K}r)}{r} + \frac{d}{dr}j_l({\cal K}r)\right]
  \,i{\bf n}\times{\bf L}Y_{lm}(\theta,\varphi)
  \label{sKV.2} \\
  {\bf L}\psi & \ = & j_l({\cal K}r)\,i{\bf L}Y_{lm}(\theta,\varphi)
  \label{sKV.3}
\end{eqnarray}

These expressions ought to be substituted now
into~\eref{totKV.a}-\eref{totKV.b}, and then into~\eref{BCondKV}
---recall that {\bf s}\,=\,${\bf s}_t\/$\,+\,${\bf s}_l\/$. It is
found thet these are equivalent to the following {\it homogeneous
linear system\/}

\begin{equation}
  \hspace*{-2.1 cm}
  \left(
  \begin{array}{ccc}
  \beta_4\left({\cal Q}R,\frac{\lambda+\gamma\lambda'}{\mu+\gamma\mu'}\right)
  & -l(l+1)\frac{{\cal K}}{{\cal Q}}\beta_1({\cal K}R) & 0 \\
  -\beta_1({\cal Q}R) & \frac{{\cal K}}{{\cal Q}}\,\beta_3({\cal K}R) & 0 \\
  0 & 0 & -\frac{{\cal K}}{{\cal Q}}\,{\cal K}R\,\beta_1({\cal K}R)
  \end{array}\right)\left(
  \begin{array}{c}
  C_0 \\ C_1 \\ C_{2}
  \end{array}\right)=0 \label{bcatlastKV}
\end{equation}
where

  \begin{eqnarray}
   \beta_0(z) & \equiv & \frac{j_l(z)}{z^2} \\
   \beta_1(z) & \equiv & \frac{d}{dz}\left[\frac{j_l(z)}{z}\right] \\
   \beta_2(z) & \equiv & \frac{d^2}{dz^2}\left[j_l(z)\right] \\
   \beta_3(z) & \equiv & \mbox{\large $\frac{1}{2}$}\,\beta_2(z) +
     \left\{\mbox{\large $\frac{l(l+1)}{2}$}-1\right\}\,\beta_0(z) \\
   \beta_4(z,A) & \equiv & \beta_2(z)-\frac{A}{2}j_l(z)
  \end{eqnarray}

The system~\eref{bcatlastKV} is to be satisfied by the constants
$C_0$, $C_1\/$ and $C_2\/$, but has no meaningful solution unless the
system matrix is {\it singular\/}, i.e., if its determinant vanishes.
Therefore

\begin{equation}
 \beta_1({\cal K}R)\,\det\left(
  \begin{array}{cc}
  \beta_4\left({\cal Q}R,\frac{\lambda+\gamma\lambda'}{\mu+\gamma\mu'}\right)
   & -l(l+1)\frac{{\cal K}}{{\cal Q}}\beta_1({\cal K}R) \\
   -\beta_1({\cal Q}R) & \frac{{\cal K}}{{\cal Q}}\beta_3({\cal K}R)
  \end{array}\right) = 0        \label{eigen}
\end{equation}

This is an equation for the parameter $\gamma\/$, on which ${\cal K}$ and
${\cal Q}$ depend through equations~\eref{KQ.a} and~\eref{KQ.b}. Clearly,
there are {\it two\/} families of solutions, or {\it eigenmodes\/},
to~\eref{eigen} associated to the vanishing of either of the two factors
in its lhs, i.e., $\beta_1({\cal K}R)$ {\it or\/} the determinant of the
displayed 2$\times$2 matrix. They are called {\it toroidal\/} and
{\it spheroidal\/} solutions, respectively.

\end{document}